\newtheorem{dfn}{Definition}%[section]
\newtheorem{thm}[dfn]{Theorem}
\newtheorem{lmma}[dfn]{Lemma}
\newtheorem{ppsn}[dfn]{Proposition}
\newtheorem{crlre}[dfn]{Corollary}
\newtheorem{xmpl}[dfn]{Example}
\newtheorem{rmrk}[dfn]{Remark}

\newcommand{\bdfn}{\begin{dfn}}
\newcommand{\bthm}{\begin{thm}}
\newcommand{\blmma}{\begin{lmma}}
\newcommand{\bppsn}{\begin{ppsn}}
\newcommand{\bcrlre}{\begin{crlre}}
\newcommand{\bxmpl}{\begin{xmpl}}
\newcommand{\brmrk}{\begin{rmrk}}

\newcommand{\edfn}{\end{dfn}}
\newcommand{\ethm}{\end{thm}}
\newcommand{\elmma}{\end{lmma}}
\newcommand{\eppsn}{\end{ppsn}}
\newcommand{\ecrlre}{\end{crlre}}
\newcommand{\exmpl}{\end{xmpl}}
\newcommand{\ermrk}{\end{rmrk}}

\def\be{\begin{equation}}
\def\ee{\end{equation}}
\def\ba{\begin{array}}
\def\ea{\end{array}}

\def\Cb{{\Bbb C}}

\documentstyle[12pt]{article}
\begin{document}
\input amssym.def
\parskip=4pt
\parindent=18pt
\baselineskip=18pt \setcounter{page}{1}
\centerline{\Large\bf
Local Invariants for a Class of Mixed States}
\vspace{4ex}
\begin{center}
Sergio Albeverio$^{a}$ \footnote{ SFB 611; BiBoS; IZKS; CERFIM(Locarno);
Acc. Arch.; USI(Mendriso)

~~~e-mail: albeverio@uni-bonn.de}, Shao-Ming Fei$^{a,b}$
\footnote{e-mail: fei@uni-bonn.de}, Debashish Goswami$^{c}$
\footnote{research partially supported by Av Humboldt
Foundation;

~~~e-mail: goswamid@isical.ac.in}

\vspace{3ex}
\begin{center}
\begin{minipage}{5.6in}

{\small $~^{a}$ Institut f\"ur Angewandte Mathematik,
Universit\"at Bonn, D-53115}

{\small $~^{b}$ Department of Mathematics, Capital Normal
University, Beijing 100037}

{\small $~^{c}$ Statistics and Mathematics Unit, Indian Statistical
Institute, Kolkata 700108}

\end{minipage}
\end{center}
\end{center}

\vskip 1 true cm
\parindent=18pt
\parskip=6pt
\begin{center}
\begin{minipage}{5in}
\vspace{3ex} \centerline{\large Abstract} \vspace{4ex}

We investigate the equivalence of quantum states under local
unitary transformations. A complete set of invariants under local
unitary transformations is presented for a class of mixed states.
It is shown that two   states in this class are locally equivalent
if and only if all these invariants have equal values for them.

\bigskip
\medskip
\bigskip
\medskip

\vfill

\end{minipage}
\end{center}

Quantum entangled states are playing very important roles in
quantum information processing and quantum computation \cite{books}.
The properties of entanglement for multipartite quantum systems
remain invariant under local unitary transformations on the subsystems.
Hence the entanglement can be characterized by all the invariants
under local unitary transformations. A complete set of invariants
gives rise to the classification of the quantum states under local
unitary transformations. Two quantum states are locally
equivalent if and only if all these invariants have equal
values for these states.
In \cite{Rains,Grassl}, a generally non-operational method has
been presented to compute all the invariants of local unitary
transformations.
In \cite{makhlin}, the invariants for general two-qubit systems are
studied and a complete set of 18 polynomial invariants is
presented. In \cite{linden} the invariants for
three qubits states are also discussed.
In \cite{generic} a complete set of invariants for generic
density matrices with full rank has been presented.

In the present paper we investigate the invariants for arbitrary
(finite-) dimensional bipartite quantum systems. We present a
complete set of invariants for a class of quantum mixed states and
show that two of these density matrices are locally equivalent if
and only if all these invariants have equal values for these
density matrices.

\section{Invariants for a class of states with arbitrary
rank}
Let us consider a general mixed state $\rho$ in a bi-partite
$n \times n$ system  $H \otimes H$ ($n \geq 2$), with a given
orthonormal basis $\{|1>,...,|n> \}$ of $H$. $\rho$ has the
eigen-decomposition
$$
\rho=\sum_{l=0}^N \mu_l |\xi_l><\xi_l|,
$$
where the rank of $\rho$ is $r(\rho)=N+1$ ($N \geq 1$), $\mu_l$ are
eigenvalues with the eigenvectors $|\xi_l>=\sum_{ij}
\xi^{(l)}_{ij}|ij>$ (and $|\xi_l><\xi_l|$ denotes, as usual, the
projector onto $|\xi_l>$), $\xi^{(l)}_{ij}\in\Cb$. Let $A_l$
denote the matrix with entries $\xi^{(l)}_{ij}$.
We call a matrix ``multiplicity free'' if each of its singular values has
multiplicity one. Let ${\cal F}$ denote the class of states $\rho$
for which $A_0$ is  multiplicity free. We shall find a complete
set of local invariants for the class ${\cal F}$, such that any pair of states belong to
${\cal F}$ are equivalent under local unitary transformations if and only if
they have the same values of these invariants.

Let $(\psi_1,...\psi_n)$, $(\eta_1,...,\eta_n)$ be orthonormal
bases such that $ A_0 =\sum_i \lambda_i |\psi_i><\eta_i|$ is the
singular value decomposition of $A_0$, where $\lambda_1 >
...>\lambda_n$ denote the singular values arranged in the
decreasing order.  Let $b^{(l)}_{ij}:=<\psi_i|A_l \eta_j>$ for
$l=1,2,...,N$, and for positive integers $k,r\geq 1$, and
multi-indices $\underline{i}=(i_1,...i_{k+1})$, (with $i_p$'s all
distinct), $\underline{j}=(j_1,...,j_{r+1})$ (with $j_q$'s all distinct),
where $i_p,j_q \in \{1,...,n \}~ \forall p,q$,
$\underline{l}=(l_1,...,l_k)$, $\underline{m}=(m_1,...,m_r)$
($l_t,m_s \in \{ 1,...,N \}$) with $i_1=j_1$, $i_{k+1}=j_{r+1}$,
and such that $(\underline{i},\underline{l}) \ne
(\underline{j},\underline{m}),$  we define
\be\label{inva}
I^\rho(\underline{i},\underline{j},\underline{l},\underline{m}):=\frac{b^{(l_1)}_{i_1i_2}
...b^{(l_k)}_{i_ki_{k+1}}}{b^{(m_1)}_{j_1j_2}...b^{(m_r)}_{j_rj_{r+1}}}
\ee
whenever the denominator in the above formula is nonzero. Let
$\Sigma^\rho$ be the set of
$(\underline{i},\underline{j},\underline{l},\underline{m})$ such
that
$I^\rho(\underline{i},\underline{j},\underline{l},\underline{m})$
is well defined.

The  following theorem is an immediate consequence of Lemma
\ref{lma1}, Lemma \ref{lma2} and the remark \ref{remark}.

\bthm
\label{thm1}
Two quantum states in ${\cal F}$ with the same rank and eigenvalues $\mu_l$, $l=0,...,N$,
are equivalent under local unitary transformations if and only if
they have the same values of the following invariants:
\be\label{inva1} \ba{l}
1)~~{\rm  Matrices}~ (B_l)_{ij}=|<\psi_i,A_l \eta_j>|,~~i,j=1,...,n,~l=1,...,N,\\[2mm]
2)~~{\rm Vector}~ C=( <\psi_1,A_0 \eta_1>,...,<\psi_n,A_0 \eta_n>),\\[2mm]
3)~~{\rm Vectors}~ D_l=(<\psi_1,A_l
\eta_1>,...<\psi_{n-1}, A_l \eta_{n-1}>),~~l=1,...,N,\\[2mm]
4)~~I^\rho~{\rm with ~the ~domain}~\Sigma^\rho.
\ea \ee \ethm
{\it Proof :} It is clear that the quantities above
are local invariant. Let
us prove that these invariants are
complete for the class ${\cal F}$. Suppose that $\rho$ and
$\rho^\prime$ are two states in the class ${\cal F}$ such that they have the
same values of these invariants. Let $\rho=\sum_{l=0}^N \mu_l
|\xi_l><\xi_l|$ and $\rho^\prime=\sum_{l=0}^N \mu_l
|\xi^\prime_l><\xi^\prime_l|$ be the eigen-decomposition of the
two states, and let $A_l=(a^{(l)}_{ij})$, $A_l^\prime=(
a^{\prime(l)}_{ij})$ be $n \times n$ complex matrices associated
with the decomposition of $\xi_l$ and $\xi^\prime_l$ respectively,
that is, $\xi_l=\sum_{ij} a^{(l)}_{ij} |ij>,$ and
$\xi^\prime_l=\sum_{ij} a^{\prime(l)}_{ij} |ij>.$ By assumption,
$A_0$ and $A_0^\prime$ are multiplicity-free, with the
singular-value decomposition
$$A_0 =\sum_i \lambda_i |\psi_i><\eta_i|,~~~~A^\prime_0
=\sum_i \lambda^\prime_i |\psi^\prime_i><\eta^\prime_i|,
$$
with the singular values arranged in the decreasing order.  Since
$\lambda_i=<\psi_i,A_0 \eta_i>$ and $\lambda_i^\prime=<\psi_i^\prime,A_0^\prime \eta_i^\prime>$,
it follows that $\lambda_i=\lambda^\prime_i$ for all
$i$. Set $(B_l)_{ij}=<\psi_i, A_l \eta_j>$, $(B_l^\prime)_{ij}=
<\psi_i^\prime,A^\prime_l \eta^\prime_j>)$ for $l=0,1,...,N$.
It is easy to see from the equalities  of
$I^\rho(\underline{i},\underline{j},\underline{l},\underline{m})$
and
$I^{\rho^\prime}(\underline{i},\underline{j},\underline{l},\underline{m})$
that the condition (III) of Lemma \ref{lma1} holds. The conditions
(I) and  (II) of Lemma \ref{lma1} also follows from the equalities
of the invariants labeled by 3) and 1) in (\ref{inva1})
respectively. Thus, by Lemma \ref{lma2} of Appendix, we conclude
that there exist unitary matrices $U$ and $V$ such that $U A_l
V^*=A^\prime_l$ for $l=0,1,...,N.$ Clearly, we have,
$\xi_i^\prime=\sum_{ij} {a^{(l)}}^\prime_{ij} |ij>=\sum_{ij}
\sum_{kl} u_{ik} a_{kl} \overline{v}_{jl}
|ij>=\sum_{kl}a_{kl}(\sum_i u_{ik}|i> ) \otimes (\sum_j
\overline{v}_{jl} |j>)=(U \otimes \bar{V})\xi_l,$ where
$U=(u_{ij}),$  $V=(v_{ij}),$ $\bar{V}=(\bar{v}_{ij}).$ Thus,
$\rho^\prime=(U \otimes \bar{V})\rho (U \otimes \bar{V})^*.$
 \hfill$\Box$

As an example we calculate all the invariants
for the Werner state \cite{werner}, $\rho_w=(1-p)I_{4\times
4}/4+p|\Psi_-><\Psi_-|$, where $0\leq p\leq 1$, $I_{4\times 4}$ is
the $4\times 4$ identity matrix and
$|\Psi_->=\frac{1}{\sqrt{2}}(|01>-|10>)$.

Here, $N=3$, $\mu_1=\mu_2=\mu_3=\frac{1-p}{4},$
$\mu_4=\frac{3p+1}{4}.$ We have,
$$\xi_0=|00>,~~\xi_1=|11>,~~~
\xi_2=\frac{1}{\sqrt{2}}(|01>+|10>),~~\xi_3=\frac{1}{\sqrt{2}}(|01>-|10>).$$
Hence $A_0=\left( \ba{cc}
1&0\\
0&0\ea \right),$ $A_1=\left( \ba{cc}
0&0\\
0&1\ea \right),$ $A_2=\left( \ba{cc}
0&\frac{1}{\sqrt{2}}\\
\frac{1}{\sqrt{2}}&0\ea \right),$ $A_3=\left( \ba{cc}
0&\frac{1}{\sqrt{2}}\\
-\frac{1}{\sqrt{2}}&0\ea \right).$ The orthonormal bases $\{
\psi_1,\psi_2 \}$ and $\{ \eta_1,\eta_2 \}$ can be chosen to be
the canonical basis $\{ |0>,|1> \}$. Thus, $(B_l)_{ij}=|<\psi_i,A_l \eta_j>|=|(A_l)_{ij}|$ in this case.
The invariants are:\\
1)$B_1=\left( \ba{cc}
0&0\\0&1\ea \right)$,
$B_2=\left( \ba{cc}
0&\frac{1}{\sqrt{2}}\\
\frac{1}{\sqrt{2}}&0\ea \right)$, $B_3=\left( \ba{cc}
0&\frac{1}{\sqrt{2}}\\
\frac{1}{\sqrt{2}}&0\ea \right)$;\\
2) $C=(1,0)$;\\
3) $D_l=0$, $l=1,2,3$;\\
4) $\Sigma^\rho=\{ ((i_1,i_2),(j_1,j_2),(l_1),(m_1));~~l_1,m_1
\in \{2,3\}, i_p,j_q \in \{ 1,2 \},~ i_1 \ne i_2, j_1 \ne j_2,~
(i_1,i_2,l_1) \ne (j_1,j_2,m_1)\},$ which can be explicitly written as:\\
$\{ \left((1,2),(1,2),(2),(3)\right);
\left((1,2),(1,2),(3),(2)\right);
\left((1,2),(2,1),(2),(2)\right); \left((1,2),(2,1),(2),(3)\right);\\
\left((1,2),(2,1),(3),(2)\right); \left((1,2),(2,1),(3),(3)\right);
\left((2,1),(1,2),(2),(2)\right); \left((2,1),(1,2),(2),(3)\right);\\
\left((2,1),(1,2),(3),(2)\right); \left((2,1),(1,2),(3),(3)\right);
\left((2,1),(2,1),(2),(3)\right); \left((2,1),(2,1),(3),(2)\right)\}$.
The values of $I^\rho$ on the above elements (in the same
order) are $1,1,1,-1,1,-1,1,1,-1,-1,-1,-1$.

\brmrk The class of states $\cal F$ for which our result works is
indeed a large one. In fact, ${\cal F }\bigcap {\cal F}_k$ is
dense (in norm) in ${\cal F}_k$, where ${\cal F}_k$ denotes the
set of $n \times n$ bipartite states of rank $k+1$, $k \geq 0.$
Consider any state $\rho \in {\cal F}_k$, with the
eigen-decomposition $\rho=\sum_{l=0}^k \mu_l |\xi_l><\xi_l|,$ with
$\xi_l=\sum_{ij} \xi^l_{ij} |ij>,$ and suppose that
$A_0:=(\xi^0_{ij})_{ij=1}^n$ is not necessarily multiplicity-free.
We claim that for any $\epsilon
>0,$ we can choose an $n \times n$ multiplicity-free matrix
$A^\prime_0=(a^\prime_{ij})$  such that $|a_{ij}-a^\prime_{ij}|
\leq n \epsilon$ $\forall~i,j$. Indeed, if $A_0=\sum_i \lambda_i
|\psi_i><\eta_i|$ is the singular value decomposition  of $A_0$,
where $\lambda_i$'s may not be all distinct, we can choose
$\lambda^\prime_i$'s which are distinct among themselves, with
$|\lambda_i-\lambda_i^\prime| \leq \epsilon$ for all $i.$
$A^\prime_0$ can be taken to be the matrix $\sum_i
\lambda_i^\prime |\psi_i><\eta_i|.$ Now, $A_0^\prime$ is
multiplicity-free, and if we choose $\rho^\prime= \mu_0
|\xi^\prime_0><\xi^\prime_0|+\sum_{l=1}^k \mu_l |\xi_l><\xi_l|,$
where $\xi^\prime_0=\sum_{ij} a^\prime_{ij} |ij>,$ it is easy to
see that $\rho^\prime \in {\cal F}_k \bigcap {\cal F}$ and $\|
\rho-\rho^\prime \| \leq 2n^3 \epsilon.$

\ermrk

\section{The invariants for another class of rank two states}

We now consider another class of states which are rank two states
on $\Cb^n \times \Cb^n$ such that  the matrices $A_0$, $A_1$ are
of
the following form :
\be \ba{l}
A_0=pP+(1-p)(1-P),~~~~
A_1=qQ+(1-q)(1-Q),
\ea \ee
where $0<p$, $q<1$ and  $P$,
$Q$  are  projection operators. We denote this class of states by $\cal G.$
\bthm

The following is a complete set of local invariants for the states in class
$\cal G$:
\be\label{inva2} \ba{l}
Tr(\rho^2),~~ Tr(A^2_0),~~ Tr(A^2_1);\\
Tr[((2P-1)(2Q-1))^k],~~~
Tr[((2P-1)E_{\pm})^k],~~~ k=1,...,n,
\ea \ee
where $E_\pm$ denotes
the projection onto the eigenspace of $(2P-1)(2Q-1)$ corresponding
to the eigenvalue $\pm 1$.
\ethm
{\it Proof :}
Clearly, the above quantities are local invariants. We show that
they are complete. Let $\rho^\prime$ be another state in $\cal G.$,
with $p^\prime$, $q^\prime$, $P^\prime$ and
$Q^\prime$ instead of $p$, $q$, $P$ and $Q$ respectively. Since
$\rho$ has two eigenvalues and $Tr(\rho)=1$, the eigenvalues are
determined by $Tr(\rho^2)$. Similarly, $Tr(A^2_0)$ and $Tr(A^2_1)$
completely determine $p$ and $q$. Thus, $p=p^\prime$ and
$q=q^\prime$. Furthermore, by Lemma \ref{lma3}, the equality of
the traces $ Tr[((2P-1)(2Q-1))^k]$,
$Tr[((2P-1)E_{\pm})^k]$, $k=1,...,n_i$,  with their primed
counterparts implies that we can find unitary matrix $U$ such that
$UPU^*=P^\prime$, $UQU^*=Q^\prime.$  This proves that $\rho$ and
$\rho^\prime$ are locally equivalent. \hfill $\Box$

This Theorem applies to a class of $d$-computable states
\cite{fei-jlw}, with a slight modification as follows. Consider a
fixed local unitary operator $W=T_1 \otimes T_2$, and let ${\cal G}_W$
denote the set of states $\rho$ such that $W \rho W^* \in {\cal
G}.$ Clearly, any two states $\rho$ and $\rho^\prime$ in ${\cal
G}_W$ are locally equivalent if and only if $W\rho W^*$ and
$W\rho^\prime W^*$ in $\cal G$ are locally equivalent too, which can be determined by
computing the invariants (\ref{inva2}).

For example, we consider a pure state on $\Cb^4 \times \Cb^4$,
$\vert\psi\rangle=\sum_{i,j=1}^4 a_{ij}|ij>$, $a_{ij}\in\Cb$,
$\sum_{i,j=1}^4 a_{ij}a_{ij}^\ast=1$. Suppose that the matrix
$A=(a_{ij})$ has the form \be\label{a} A=\left( \ba{cccc}
0&0&a_1&b_1\\
0&0&\bar{b}_1&d_1\\
a_1& b_1&0&0\\
\bar{b}_1&d_1&0&0 \ea \right),
\ee
$a_1,b_1,d_1 \in\Cb$,
satisfying $a_1,d_1 \geq 0,$ $a_1d_1 \geq |b_1|^2.$ In this case,
$\vert\psi\rangle$ is a $d$-computable state and its entanglement
of formation is a monotonically increasing function of the
generalized concurrence $d=4(a_1d_1-|b_1|^2)$. The entanglement of
formation for any mixed states with decompositions on
$d$-computable states can be calculated analytically. Let
$\vert\psi^\prime>=\sum_{i,j=1}^4 a_{ij}^\prime|ij>$ be another
pure state with $A^\prime=(a_{ij}^\prime)$ of the form (\ref{a})
and $<\psi^\prime|\psi>=0$. Then
\be\label{example} \rho=\mu
\vert\psi><\psi| + (1-\mu) \vert\psi^\prime><\psi^\prime|
\ee
is an entangled rank two density matrix. Set $T=\left( \ba{cc}
0&I_2\\
I_2&0\\ \ea \right)$ and $W=T \otimes I_4.$  As the matrices $A$,
$A^\prime $ are of the form $TB$, where $B$ is a nonnegative
matrix with  at most two different eigenvalues with degeneracy
two, $\rho \in {\cal G}_W$, and  the invariants (\ref{inva2})
determine the equivalence of two mixed states of the form
(\ref{example}) under local unitary transformations.

\section{Remarks and conclusions}

We have investigated the equivalence of quantum bipartite states
under local unitary transformations. For the states $\rho$ for
which $A_0$  is multiplicity free, as well as for the states
$\rho$ which are of rank two on $\Cb^n \times \Cb^n$ such that
each of the matrices $A_0$ and $A_1$ is  a nonnegative matrix
having at most two different eigenvalues, a complete set of
invariants under local unitary transformations is presented. Two
of these states are locally equivalent if and only if all these
invariants have equal values for them.

The results can be generalized to the multipartite case. For
instance, we can consider a tripartite state $\rho_{ABC}$ with
subsystems, say, $A,B$ and $C$ as bipartite states $\rho_{A|BC}$,
$\rho_{AB|C}$ or $\rho_{AC|B}$. If the conditions in our theorems
are satisfied for one of the bipartite decompositions, say
$\rho_{A|BC}$, we can judge whether two such tripartite states are
equivalent or not under local unitary transformations, in this
bipartite decomposition. If they are, we consider further
$\rho_{BC}=Tr_A (\rho_{A|BC})$, which is again a bipartite state
and can be judged by using our theorems, if the related conditions
are satisfied. In this way the equivalence for a class of
multipartite states can also be studied according to our theorems.

\section{APPENDIX}
\blmma \label{lma1*} Let $B_l=( b^{(l)}_{ij} ),$ $C_l=(
c^{(l)}_{ij} )$ be $n \times n$  matrices with complex entries,
$l=1,...,N$, where $n$ and $N$ are positive integers. Then there
exist complex numbers $u_i,i=1,...,n$, with $|u_i|=1$, $\forall i$
and $c^{(l)}_{ij}=\frac{u_i}{u_j}b^{(l)}_{ij}$ for all
$i,j=1,...,n$,
$l=1,...,N$ if and only if the following conditions hold :\\
(I) $b^{(l)}_{ii}=c^{(l)}_{ii}$ $\forall i,l$,\\
(II) $|b^{(l)}_{ij}|=|c^{(l)}_{ij}|$ $\forall i,j,l,$\\
(III) For all choices of $l_1,...,l_k,m_1,...,m_r \in \{1,2,...,N
\}$ ($k,r \geq 1$), $i_1,...,i_{k+1}, j_1,...,j_{r+1} \in
\{1,2,...,n \}$ with $i_1=j_1, i_{k+1}=j_{r+1},$
$$
b^{(l_1)}_{i_1i_2}b^{(l_2)}_{i_2i_3}...b^{(l_k)}_{i_k
i_{k+1}}c^{(m_1)}_{j_1j_2}...c^{(m_r)}_{j_rj_{r+1}}=
c^{(l_1)}_{i_1i_2}c^{(l_2)}_{i_2i_3}...c^{(l_k)}_{i_ki_{k+1}}b^{(m_1)}_{j_1j_2}...
b^{(m_r)}_{j_rj_{r+1}}.$$ \elmma
{\it Proof :}
The proof of the necessity of the conditions (I), (II), (III) is
trivial. We prove the sufficiency of these conditions. Assume that
(I), (II), (III) are satisfied. We define a relation $\sim$ on the
set $\{ 1,2,...,n \}$ as follows. Let us set $i \sim i$ for all
$i$, and
 for $i,j$ different, let us say $i \rightarrow j$ if   there exist
  $i_1,...,i_{k+1}$ ($k \geq 1$) with $i_1=i, i_{k+1}=j$ and $l_1,...,l_k$ such that
  $b^{(l_1)}_{i_1i_2},b^{(l_2)}_{i_2i_3},...,b^{(l_k)}_{i_ki_{k+1}}$ are
  all nonzero (by (II) this is equivalent to saying that similar quantities with $b$ replaced
   by $c$ are all nonzero). We set $i \sim j$ (for different $i,j$) if $i \rightarrow j$ and $j \rightarrow i.$
   It is easy to verify that $\sim$ is an equivalence
  relation.  Let $\{
   1,2,...,n \}=E_1 \bigcup ... \bigcup E_p$ ($p \geq 1$) be the
   decomposition into equivalence classes. Choose and fix any
   $i_1^*,...,i_p^*$ from $E_1,...,E_p$ respectively. Set $u_i=1$
   for $i \in \{ i_1^*,...,i_p^* \}$. For any other $i$, say $i
   \in E_t$ ($1 \leq t \leq p$), but $i \neq i_t^*,$ we define
   $$
   u_i := \frac{c^{(l_1)}_{i_1i_2}...c^{(l_k)}_{i_ki_{k+1}}}
   {b^{(l_1)}_{i_1i_2}...b^{(l_k)}_{i_ki_{k+1}}},
   $$
    where $i_1=i, i_2,...,i_k,i_{k+1}=i_t^*$, $l_1,...,l_k$ are chosen
    such that $ b^{(l_1)}_{i_1i_2},...,b^{(l_k)}_{i_ki_{k+1}}$ are
    nonzero, which exist as $i \sim i_t^*.$ $u_i$ is well
    defined by (III). Indeed, if any other such ``path''
    $i_1^\prime=i,i_2^\prime,...,i_{r+1}^\prime=i_t^*$,
    $l^\prime_1,...,l^\prime_r$ is used, we have by (III)
     $$\frac{c^{(l_1)}_{i_1i_2}...c^{(l_k)}_{i_ki_{k+1}}}
   {b^{(l_1)}_{i_1i_2}...b^{(l_k)}_{i_ki_{k+1}}}=
   \frac{c^{(l^\prime_1)}_{i^\prime_1i^\prime_2}...c^{(l^\prime_r)}_{i^\prime_ri^\prime_{r+1}}}
   {b^{(l^\prime_1)}_{i^\prime_1i^\prime_2}
   ...b^{(l^\prime_r)}_{i^\prime_ri^\prime_{r+1}}},$$
    which shows that $u_i$ remains the same if the primed sequence
    is used. Note also that by (II), we have $|u_i|=1$ for all $i.$

With this definition of the $u_i,$'s we claim that \be\label{star}
c^{(l)}_{ij}=\frac{u_i}{u_j}b^{(l)}_{ij} \ee for all $l,i,j$. For
$i=j$, (\ref{star}) follows from (I). In case $b^{(l)}_{ij}=0$,
the relation (\ref{star}) follows from (II). The only nontrivial
case to prove arises when $b^{(l)}_{ij}$ (and hence also
$c^{(l)}_{ij}$) is nonzero for $i \neq j$. Thus, $i,j$ can be
assumed to belong to the same equivalence class, say $E_t$. If
$j=i^*_t,$ we can take $k=1$, with $i_1=i$, $i_2=j$ in the
definition of $u_i$, and the relation $(\ref{star})$ follows.
Otherwise, i.e. if $j \neq i_t^*,$ we choose sequences $i_1=i$,
$i_2,...,i_{k+1}=i_t^*$, $l_1,...,l_k$ for
 the definition of $u_i$, and $j_1=j$, $j_2,...,j_{r+1}=i_t^*$,
 $m_1,...,m_r$ for the definition of $u_j$, so that
 $$
 \frac {u_i}{u_j} \frac{b^{(l)}_{ij}}{c^{(l)}_{ij}}=
  \frac{ b^{(l)}_{ij}b^{(m_1)}_{jj_2}...b^{(m_r)}_{j_ri^*_t}}{
  c^{(l)}_{ij}c^{(m_1)}_{jj_2}...c^{(m_r)}_{j_ri^*_t}}.
  \frac{c^{(l_1)}_{ii_2}...c^{(l_k)}_{i_ki^*_t}}{b^{(l_1)}_{ii_2}...b^{(l_k)}_{i_ki^*_t}}=1
  $$
   by (III).
   This completes the proof of the Lemma. \hfill $\Box$
\brmrk
\label{remark}
In the statement of the above Lemma, it is easy to see that in the
condition (III) it is enough to consider distinct
$i_1,i_2,...,i_{k+1}$ and distinct $j_1,...j_{r+1}$, as
$b^{(l)}_{ii}=c^{(l)}_{ii}$ for all $i,l.$
\ermrk
Now we state and prove a result which is a slight variation of
Lemma \ref{lma1*}, which suits our purpose.
\blmma \label{lma1} Let $B_l=( b^{(l)}_{ij} ),$ $C_l=(
c^{(l)}_{ij} )$ be as in Lemma \ref{lma1*}, with $n \geq 2.$  Then
there exist complex numbers $u_i,i=1,...,n$, $v_n$ with $|u_i|=1$,
$\forall i,$ $|v_n|=1,$  such that
$c^{(l)}_{ij}=\frac{u_i}{u_j}b^{(l)}_{ij},$
$c^{(l)}_{in}=\frac{u_i}{v_n} b^{(l)}_{in},$
$c^{(l)}_{nj}=\frac{u_n}{u_j} b^{(l)}_{nj}$ for all
$i,j=1,...,n-1$ and
$l=1,...,N$ if and only if the following conditions hold :\\
(I) $b^{(l)}_{ii}=c^{(l)}_{ii}$ $\forall i=1,...,n-1;$ $l=1,...,N$;\\
(II) $|b^{(l)}_{ij}|=|c^{(l)}_{ij}|$ $\forall i,j=1,...,n$, $l=1,...,N;$\\
(III) For all choices of $l_1,...,l_k,m_1,...,m_r \in \{1,2,...,N
\}$ ($k,r \geq 1$), $i_1,...,i_{k+1}, j_1,...,j_{r+1} \in
\{1,2,...,n \}$ with $i_1=j_1, i_{k+1}=j_{r+1},$ and with the
restriction that $(i_1,...,i_{k+1})$ are all distinct and so are
$(j_1,...,j_{r+1}),$  one has
$$
b^{(l_1)}_{i_1i_2}b^{(l_2)}_{i_2i_3}...b^{(l_k)}_{i_k
i_{k+1}}c^{(m_1)}_{j_1j_2}...c^{(m_r)}_{j_rj_{r+1}}=
c^{(l_1)}_{i_1i_2}c^{(l_2)}_{i_2i_3}...c^{(l_k)}_{i_ki_{k+1}}b^{(m_1)}_{j_1j_2}...
b^{(m_r)}_{j_rj_{r+1}}.$$
\elmma

Let $N,n \geq 1$ be  positive integers, and $A_0,A_1,...,A_N$;
$A^\prime_0,A^\prime_1,...,A^\prime_N$ be $n \times n$ positive
matrices, $n \geq 2$. Let $(\lambda_1,...,\lambda_n)$ be the
singular values of $A_0$, and
$(\lambda^\prime_1,...,\lambda^\prime_n)$ be those of
$A^\prime_0$. Assume furthermore that $(\lambda_1,...,\lambda_n)$
are all distinct, say, $\lambda_1>...>\lambda_n$, and similarly
$\lambda^\prime_1
>...>\lambda^\prime_n.$ Let $(\psi_1,....\psi_n)$,
$(\eta_1,...,\eta_n)$ be two  orthonormal bases for $C^n$ such
that the singular value decomposition of $A_0$ is given by $$A_0=
\sum_i \lambda_i |\psi_i><\eta_i|.$$
Similarly, let
$(\psi^\prime_1,...,\psi^\prime_n)$ and
$(\eta^\prime_1,...,\eta^\prime_n)$  are the orthonormal bases
  corresponding to the singular value decomposition of $A^\prime_0.$ Let matrices $B_l$, $C_l$,
$l=1,...,N$ be defined by $(B_l)_{ij}=b^{(l)}_{ij}$, $(C_l)_{ij}=
c^{(l)}_{ij}$, where $b^{(l)}_{ij}=<\psi_i, A_l \eta_j>$,
$c^{(l)}_{ij}=<\psi^\prime_i,A^\prime_l \eta^\prime_j>$. We have:
\blmma \label {lma2}
There exist two unitary matrices $U,V$ such that $UA_l
V^*=A^\prime_l$ for all $l=0,1,...,N$ if and only if
$\lambda_i=\lambda^\prime_i$ $\forall i$ and the conditions (I),
(II) and (III) in the statement of Lemma \ref{lma1} are satisfied
for the choices of $b^{(l)}_{ij}$, $c^{(l)}_{ij}$'s as above.
\elmma
{\it Proof :}
Let $V_1,V_2,V_1^\prime, V_2^\prime $ be unitary matrices such
that   $V_1A_0V_2^*=D_0:=diag(\lambda_1,...\lambda_n)$ and
$V_1^\prime
A^\prime_0{V_2^\prime}^*=D^\prime_0:=diag(\lambda^\prime_1,...\lambda^\prime_n).$
Clearly, $V_1A_lV_2^*=B_l$, $V_1^\prime
A^\prime_l{V_2^\prime}^*=C_l$ for $l=1,...,N.$

{\it Proof of the ``if'' part :}
Here, $D_0=D^\prime_0=D$, say. By Lemma \ref{lma1}, we can find
$u_i,i=1,...,n,$ $v_n$ with $|u_i|=1,$ $|v_n|=1,$ and
$c^{(l)}_{ij}=\frac{u_i}{v_j}b^{(l)}_{ij}$ $\forall i,j=1,...,n$,
$l=0,...,N$,
 with $v_j=u_j$ for $j=1,...,n-1.$    In other words, $C_l=W_1B_lW_2^*$, $l=1,...,N$,
where $W_l$ is the unitary given by $W_l:=diag(u_1,...u_n)$ and
similarly, $W_2:=diag(u_1,...,u_{n-1},v_n).$ We take
$U:={V_1^\prime}^*W_1V_1,$ $V={V_2^\prime}^*W_2V_2,$ and it is
easy to verify that $UA_lV^*=A^\prime_l$ for $l=0,1,...,N.$

{\it Proof of the ``only if '' part :}
Suppose now that there are  unitary matrices $U,V$ such that
$UA_lV^*=A^\prime_l$ for $l=0,1,...,N$. It follows from the
assumption $UA_0V^*=A^\prime_0$ that $D_0=D^\prime_0=D$, say. We
have $UA_0V^*=UV_1^*D V_2
V^*={V_1^\prime}^*DV_2^\prime=A_0^\prime,$ from which it follows
that $W_1 D=D W_2,$ where $W_1=V_1^\prime U V_1^*,$
$W_2=V_2^\prime V V_2^*.$  Thus, $W_1DD^*
W_1^*=DW_2^*W_2D^*=DD^*.$ Since $D$ is diagonal with all entries
distinct and nonnegative,
$DD^*=D^2=diag(\lambda_1^2,...,\lambda_n^2).$ It follows that
$W_1$ must  also be diagonal, i.e. $W_1=diag(u_1,...,u_n)$ for
some $u_1,...,u_n$ with $|u_i|=1$. Similarly, $W_2$ is diagonal,
say $diag(v_1,...,v_n).$ Furthermore, we have $W_1D=DW_2$, which
implies that $\lambda_i u_i= \lambda_i v_i$ for all $i$, and as
$\lambda_1,...,\lambda_{n-1}$ are strictly positive numbers (only
$\lambda_n$ can possibly be $0$), we conclude that $u_i=v_i$ for
$i=1,...,n-1.$  Obviously, $C_l=W_1B_lW_2^*$, from which the
conditions (I), (II) and (III) of Lemma \ref{lma1} follow. \hfill
$\Box$
\blmma
\label{lma3}
Let $(P,Q)$ and $(P^\prime, Q^\prime)$ be two pairs of projections
in $n$-dimensional ($n \geq 1$) Hilbert space. There exits a
unitary matrix $U$ such that $P^\prime=UPU^*$ and $Q^\prime=UQU^*$
if and only if the following conditions are satisfied:
\be\label{cond1} \ba{l} (I)~~~ Tr[((2P-1)(2Q-1))^m]=
Tr[((2P^\prime-1)(2Q^\prime-1))^m],~m=1,...,n;\\
(II)~~~ Tr[((2P-1)E_{\pm})^m]=
Tr[((2P^\prime-1)E^\prime_{\pm})^m],~m=1,...,n, \ea \ee where
$E_+$ and $E_-$ denote the projection onto the eigenspace of the
eigenvalue $1$ and $-1$ of the unitary matrix $(2P-1)(2Q-1)$
respectively. $E^\prime_\pm$ are defined similarly, replacing $P$
and $Q$ by $P^\prime$ and $Q^\prime$.
 \elmma
{\it Proof :}
The result can be proved by applying the characterization of a
pair of projections obtained by Halmos \cite{Hal} (see also
\cite{Par} and the references therein for related discussion). We,
however, present a direct proof in our finite-dimensional
situation.

The ``only if" part is trivial. So we suppose that the conditions
(I) and (II) hold. Let $S=2P-1$, $V=(2P-1)(2Q-1)$, and
$S^\prime=2P^\prime-1$, $V^\prime=(2P^\prime-1)(2Q^\prime-1)$. $S$
and $S^\prime$ are selfadjoint unitary matrices, $V$ and
$V^\prime$ are unitary ones. We also have $SVS=V^*$, $S^\prime
V^\prime S^\prime={V^\prime}^*$. Note that by (I), the eigenvalues of $V$
and $V^\prime$ are the same, and have the same multiplicities. Let
$\Delta$ be the set of these eigenvalues, and $\Delta_+$ (resp.
$\Delta_-$) be the set of eigenvalues with positive ( resp.
negative) imaginary parts. Furthermore, if we denote by
$H_\lambda$ (resp. $H^\prime_\lambda$) the eigenspace of $V$
(resp. of $V^\prime$) corresponding to the eigenvalue $\lambda$
($dim (H_\lambda)=dim(H^\prime_\lambda)$, as is already noted),
then it is easy to verify that $SH_\lambda=H_{\lambda^{-1}}$, and
a similar fact is true for $S^\prime$ and $H^\prime_\lambda$. We
want to define a unitary $U$ from $\Cb^n =\oplus_\lambda H_\lambda
$ to $\Cb^n=\oplus_\lambda H^\prime_\lambda$ such that
$U=\oplus_\lambda U_\lambda$, where $U_\lambda : H_\lambda
\longrightarrow H^\prime_\lambda$ for all $\lambda$ and
$USU^*=S^\prime.$ For $\lambda \in \Delta_+,$ choose any unitary
$U_\lambda$ from $H_\lambda$ onto $H_\lambda^\prime$ (this is
possible as $H_\lambda$ and $H_\lambda^\prime$ have the same
dimension), and then for $\lambda \in \Delta_-,$ i.e.
$\lambda^{-1} \in \Delta_+,$ choose
$U_\lambda=S^\prime|_{H^\prime_{\lambda^{-1}}}= U_{\lambda^{-1}}
S|_{H_\lambda}.$ Finally, we need to define $U_{\pm 1}$, for which
we shall make use of (II). By (II), $ S|_{H_{+1}}=SE_+$ is
unitarily equivalent to $S^\prime E^\prime_+,$ so there exists a
unitary $U_{+1}$ satisfying $U_{+1}S|_{H_{+1}}U_{+1}^*=S^\prime
E^\prime_+.$ Similarly, $U_{-1}$ can be defined. By construction,
it is clear that $USU^*=S^\prime$ and $UVU^\prime=V^\prime,$ which
is equivalent to having $UPU^*=P^\prime$ and $UQU^*=Q^\prime.$
\hfill $\Box$

\end{document}